\documentclass[dvips]{article}
\usepackage{supertabular,lscape,epsfig}
\usepackage{amssymb}
\usepackage{amsmath}

\usepackage[polish]{babel}
\usepackage[T1]{fontenc}
\usepackage[latin2]{inputenc}
\usepackage{pslatex}

\DeclareSymbolFont{ppa}{OT1}{ppl}{m}{it}
\DeclareMathSymbol{\vv}{\mathalpha}{ppa}{'166}

\begin{document}

\newcommand{\dd}{\,{\rm d}}
\newcommand{\ie}{{\it i.e.},\,}
\newcommand{\etal}{{\it et al.\ }}
\newcommand{\eg}{{\it e.g.},\,}
\newcommand{\cf}{{\it cf.\ }}
\newcommand{\vs}{{\it vs.\ }}
\newcommand{\zdot}{\makebox[0pt][l]{.}}
\newcommand{\up}[1]{\ifmmode^{\rm #1}\else$^{\rm #1}$\fi}
\newcommand{\dn}[1]{\ifmmode_{\rm #1}\else$_{\rm #1}$\fi}
\newcommand{\upd}{\up{d}}
\newcommand{\uph}{\up{h}}
\newcommand{\upm}{\up{m}}  
\newcommand{\ups}{\up{s}}
\newcommand{\arcd}{\ifmmode^{\circ}\else$^{\circ}$\fi}
\newcommand{\arcm}{\ifmmode{'}\else$'$\fi}
\newcommand{\arcs}{\ifmmode{''}\else$''$\fi}
\newcommand{\MS}{{\rm M}\ifmmode_{\odot}\else$_{\odot}$\fi}
\newcommand{\RS}{{\rm R}\ifmmode_{\odot}\else$_{\odot}$\fi}
\newcommand{\LS}{{\rm L}\ifmmode_{\odot}\else$_{\odot}$\fi}

\newcommand{\Abstract}[2]{{\footnotesize\begin{center}ABSTRACT\end{center}
\vspace{1mm}\par#1\par   
\noindent
{~}{\it #2}}}

\newcommand{\TabCap}[2]{\begin{center}\parbox[t]{#1}{\begin{center}
  \small {\spaceskip 2pt plus 1pt minus 1pt T a b l e}
  \refstepcounter{table}\thetable \\[2mm]
  \footnotesize #2 \end{center}}\end{center}}

\newcommand{\TableSep}[2]{\begin{table}[p]\vspace{#1}
\TabCap{#2}\end{table}}

\newcommand{\FigCap}[1]{\footnotesize\par\noindent Fig.\  %
  \refstepcounter{figure}\thefigure. #1\par}

\newcommand{\TableFont}{\footnotesize}
\newcommand{\TableFontIt}{\ttit}
\newcommand{\SetTableFont}[1]{\renewcommand{\TableFont}{#1}}

\newcommand{\MakeTable}[4]{\begin{table}[htb]\TabCap{#2}{#3}
  \begin{center} \TableFont \begin{tabular}{#1} #4
  \end{tabular}\end{center}\end{table}}

\newcommand{\MakeTableSep}[4]{\begin{table}[p]\TabCap{#2}{#3}
  \begin{center} \TableFont \begin{tabular}{#1} #4
  \end{tabular}\end{center}\end{table}}

\newenvironment{references}%
{
\footnotesize \frenchspacing
\renewcommand{\thesection}{}
\renewcommand{\in}{{\rm in }}
\renewcommand{\AA}{Astron.\ Astrophys.}
\newcommand{\AAS}{Astron.~Astrophys.~Suppl.~Ser.}
\newcommand{\ApJ}{Astrophys.\ J.}
\newcommand{\ApJS}{Astrophys.\ J.~Suppl.~Ser.}
\newcommand{\ApJL}{Astrophys.\ J.~Letters}
\newcommand{\AJ}{Astron.\ J.}
\newcommand{\IBVS}{IBVS}
\newcommand{\PASP}{P.A.S.P.}
\newcommand{\Acta}{Acta Astron.}
\newcommand{\MNRAS}{MNRAS}
\renewcommand{\and}{{\rm and }}
\section{{\rm REFERENCES}}
\sloppy \hyphenpenalty10000
\begin{list}{}{\leftmargin1cm\listparindent-1cm
\itemindent\listparindent\parsep0pt\itemsep0pt}}%
{\end{list}\vspace{2mm}}
 
\def\TYLDA{~}
\newlength{\DW}
\settowidth{\DW}{0}
\newcommand{\dw}{\hspace{\DW}}

\newcommand{\refitem}[5]{\item[]{#1} #2%
\def\REFARG{#3}\ifx\REFARG\TYLDA\else, {\it#3}\fi
\def\REFARG{#4}\ifx\REFARG\TYLDA\else, {\bf#4}\fi
\def\REFARG{#5}\ifx\REFARG\TYLDA\else, {#5}\fi.}

\newcommand{\Section}[1]{\section{#1}}
\newcommand{\Subsection}[1]{\subsection{#1}}
\newcommand{\Acknow}[1]{\par\vspace{5mm}{\bf Acknowledgements.} #1}
\pagestyle{myheadings}

\newfont{\bb}{ptmbi8t at 12pt}
\newcommand{\xrule}{\rule{0pt}{2.5ex}}  
\newcommand{\xxrule}{\rule[-1.8ex]{0pt}{4.5ex}}  
\def\thefootnote{\fnsymbol{footnote}}
\newcommand{\uprule}{\rule{0pt}{2.5ex}}
\newcommand{\douprule}{\rule[-2ex]{0pt}{4.5ex}}
\newcommand{\dorule}{\rule[-2ex]{0pt}{2ex}}

\begin{center}
{\Large\bf
Age--Rotation--Activity Relations for M Dwarf Stars\\ 
\vskip2pt
Based on ASAS Photometric Data}
\vskip1.7cm
{\bf 
M.~~ K~i~r~a~g~a~~ and~~ K.~~ S~t~ę~p~i~e~\'n}
\vskip7mm
{Warsaw University Observatory, Al.~Ujazdowskie~4, 00-478~Warszawa, Poland\\
e-mail:(kiraga,kst)@astrouw.edu.pl}
\end{center}

\Abstract{Based on analysis of photometric observations of nearby M type
stars obtained with ASAS, 31 periodic variables were detected. The
determined periods are assumed to be related to rotation periods of the
investigated stars. Among them 10 new variables with periods longer than
10 days were found, which brings the total number of slowly rotating M
stars with known rotation periods to 12 objects.

X-ray activity and rotation evolution of M stars follows the trends
observed in G--K type stars. Rapidly rotating stars are very active and
activity decreases with increasing rotation period but the period-activity
relation is mass-dependent which suggests that the rotation period alone is
not a proper measure of activity. The investigated stars were grouped
according to their mass and the empirical turnover time was determined for
each group. It increases with decreasing mass more steeply than for K type
stars for which a flat dependence had been found. The resulting Rossby
number-activity relation shows an exponential decrease of activity with
increasing Rossby number.

The analysis of space motions of 27 single stars showed that all rapidly
rotating and a few slowly rotating stars belong to young disk (YD) whereas
all old disk (OD) stars are slowly rotating. The median rotation period of
YD stars is about 2 days and that of OD stars is equal to 47 days, \ie
nearly 25 times longer. The average X-ray flux of OD stars is about 1.7 dex
lower than YD stars in a good agreement with the derived Rossby
number-activity formula supplemented with rotation-age relation and in a
fair agreement with recent observations but in a disagreement with the
Skumanich formula supplemented with the activity-rotation relation.}{Stars:
activity -- Stars: rotation -- Stars: low-mass, brown dwarfs}

\Section{Introduction}
Chromospheric-coronal activity is inherent to lower main sequence (MS)
stars. In rotating stars with subphotospheric convection zones
magnetohydrodynamic dynamo operates generating magnetic fields inside
the zone and/or at the interface between the convection layer and the
radiative core. After emerging above the photosphere, the magnetic field
drives the heating of a stellar atmosphere and produces various activity
phenomena like spots, plages, flares and magnetized winds. Empirical
data indicate that the activity level increases with increasing rotation
rate (Hartmann and Noyes 1987, Maggio \etal 1987, Stępie\'n 1989, 1994,
Hempelmann \etal 1995). The observed period-activity relation for MS
stars from a narrow spectral type shows a small scatter but when stars
of all spectral types are plotted together, the scatter increases
substantially (Noyes \etal 1984, Stępie\'n 1994, Pizzolato \etal 2003).
This clearly shows that period-activity relation varies with stellar
mass. To decrease the scatter, rotation periods of stars of a given
spectral type must be scaled down by a mass dependent quantity called
the convective turnover time $\tau_c$. Such a scaling is suggested by a
simple parametric dynamo theory (Durney and Latour 1978, Gray 1982) in
which the Rossby number $Ro=P_{\rm rot}/\tau_c$ rather than rotation
period is a primary parameter controlling the efficiency of field
generation.

Values of turnover times for stars of various masses can be determined
from theoretical models (\eg Gilman 1980, Gilliland 1985, Rucinski and
VandenBerg 1986, Kim and Demarque 1996) or empirically (Noyes \etal
1984, Stępie\'n 1994, 2003, Pizzolato \etal 2003). The values of
$\tau_c$ for G--K type stars, given by various authors, do not differ
much but for M type stars they diverge badly.
Theoretical models indicate a steep increase of $\tau_c$ with 
decreasing mass. For example, the theoretical value of turnover time for
a 0.5~\MS\ star is 3--3.5 times longer than for a 1~\MS\ star (Kim and
Demarque 1996, Rucinski and VandenBerg 1986 -- in the latter case a
slight extrapolation is needed because their models reach only 0.7~\MS).
Purely empirical determinations indicate that $\tau_c$ increases with
decreasing mass down to about 0.8~\MS, but then it levels off till about
0.6~\MS\ (Stępie\'n 1994, 2003, Pizzolato \etal 2003). Due to a shortage
of known periods for less massive, slowly rotating stars, the behavior
of the empirical turnover time with decreasing mass beyond 0.6~\MS\ is
very uncertain. There are indications that it may start increasing again
but the situation has been obscure. The results of a recent analysis of
angular momentum evolution of very low mass stars (VLMS), based on
observations of young clusters, suggest that very long turnover times,
predicted by theoretical models, do not agree with the observed rotation
velocities of such stars. Sills, Pinsonneault and Terndrup (2000) noted
that the relation $\omega_{\rm crit}(\tau_c)$, where $\omega_{\rm crit}$
is a critical angular velocity for the occurrence of saturation, breaks
down if values of $\tau_c$ for low mass stars are extrapolated from the
higher mass models of Kim and Demarque (1996). Lower values are in a
better agreement with observations.

M type stars seem to share all characteristic properties of G--K type
active stars: the most active stars have $L_X/L_{\rm bol}$ at the level
of $10^{-3}$ \ie close to the upper limit for all active stars whereas
the lowest measured X-ray fluxes are some 3 orders of magnitude lower as
can be seen in the NEXXUS database (Schmitt and Liefke 2004). High
activity level is related to stellar youth and it decreases with age so
that old stars show notably decreased level of activity, both
chromospheric (Delfosse \etal 1998, Silvestri, Hawley and Oswalt 2005)
and coronal (Fleming, Schmitt and Giampapa 1995, Feigelson \etal 2004).
The most active M stars rotate rapidly, with periods from a fraction of
a day up to several days. Quite a number of rotation periods of such
stars is known (Pizzolato \etal 2003 and references therein). Low
activity stars seem to rotate slowly -- they show values of $\vv\sin i$
below the present resolution threshold of about 2--3~km/s (Delfosse
\etal 1998), which corresponds to a lower limit of several days for the
rotation period. This is not a very restrictive condition because the
lowest mass stars with such rotation periods are still in a saturation
regime (Pizzolato \etal 2003). To extend the period-activity relation to
the least active M type stars we need direct measurements of rotation
periods of slowly rotating stars with periods several times longer than
the saturation limit.

A standard method of measuring rotational periods is based on 
observations of stellar surface inhomogeneities which produce signal
modulations over the rotational period. Several periods of
chromospherically active stars have been determined in this way from
rotational variations of CaII emission, carried out within the Mount
Wilson program (Noyes \etal 1984, Donahue, Saar and Baliunas 1996 and
references therein). The technique works well for F--K type stars. In M
type stars the emission in CaII lines becomes a poor measure of the
chromospheric energy losses and is difficult to measure. So far, the
only M type star for which a rotation period has been determined from
calcium emission modulation, is GJ 411 (=HD 95735) with spectral type
M2V (Noyes \etal 1984). Star-spots produce another type of surface
inhomogeneities, observable with a broadband photometry. A rotational
modulation of stellar brightness has been observed in many active stars,
including those of M type (see Messina \etal 2003 for a recent review).
High activity stars have typical amplitudes of several percent which is
easy to detect, particularly for short rotation periods. Low activity
stars are expected to have much less spotedness producing
correspondingly weaker light variations and on a much longer time scale 
(Messina \etal 2003). In addition, possible variations of a spot pattern
on a time scale comparable to the rotation period can mask rotational
modulation. Long time series of observations are needed to filter out
the correct value of a period. Because of all these problems, only very
few detections of rotation periods of slowly rotating stars have been
reported from the observations of star-spots.

In the last decade many very low mass stars (VLMS) and brown dwarfs have
been detected using high sensitivity observational techniques in red and
infrared. Several studies followed, including searches for rotation
periods and measurements of $\vv\sin i$. Most of the studies are
concentrated on young clusters (Scholz and Eisl\"offel 2005 and
references therein) but several field objects have also been observed.
The results show that many observed VLMS are active and fast rotating. 
The transition from partly to fully convective stars has no apparent
influence on their activity and/or rotation (Delfosse \etal 1998, Mokler
and Stelzer 2002). The average value of $\vv\sin i$ apparently increases
with the advancing spectral type from early M to late M and L (Delfosse
\etal 1998, Mohanty and Basri 2003, Scholz and Eisl\"offel 2005). This
is usually interpreted in terms of increasing time scale for spin down
with decreasing mass but a detailed age--activity--rotation relation for
M type stars of various masses is unknown. One should not forget,
however, when interpreting observations of stars with ages of about
1~Gyr or less, that evolutionary effects should also be taken into
account due to a rapid increase of the approach time to ZAMS of VLMS, up
to 3~Gyr for least massive stars (Baraffe \etal 1998). More stars with
known age, activity level and rotation period (particularly those
inactive and/or slowly rotating) will help in solving the problem of
spin down and activity decrease of low mass stars.

The present investigation is aimed at enlargement of the number of M
type stars with known rotation periods. A particular care is taken to
detect slowly rotating stars. To this purpose an extensive data set of
photometric observations obtained during the automated sky survey ASAS
(Pojma\'nski 1997, 2004) is used. To assure a minimum reasonable
accuracy, we analyzed M stars brighter than a visual magnitude
$V=12.5$~mag which restricted our sample to M0--M3.5 type stars only,
with an exception of Proxima (M5.5).

Section~2 describes the star selection criteria and method of
analysis. In Section~3 the results of the period search are presented
and interpreted in terms of empirical turnover time and
age--activity--rotation relations. Section~4 contains the summary of the
main conclusions. 

\Section{Star Selection and Analysis}
ASAS cameras are located at Las Campanas Observatory, Chile, and the
observations cover the sky south of declination +28\arcd. At the angular
resolution of 14\arcs/pixel several visual binaries are not separated.

The most accurate ASAS photometric measurements are obtained for stars
with $V$ magnitudes between 8.0 and 12.5. Brighter stars are usually
saturated, whereas the observational errors increase rapidly for fainter
stars. For the period search we selected stars with the following
properties:

(i) X-ray flux is known from H\"unsch \etal (1999) or from NEXXUS
database

(ii) spectral type is M, 

(iii) $8<\bar V<12.5$, where $\bar V$ is a mean $V$ magnitude.

180 stars from the ASAS database fulfilled these conditions. Only good
quality photometric data (with quality grade ``A'' and ``B'' -- Pojma\'nski
2006, private communication) were included into the period analysis.

Objects monitored by ASAS are typically observed once per night or,
sometimes, even more sparsely, depending on weather. The Nyiquist
frequency corresponds in this case to the period of 2 days. We looked
for periods from 0.2 d up to the length of an observational run in a
single season ($\approx100$ days). We assumed that season to season
variations are not connected with rotation and we eliminated them before
performing period search. Because the shortest periods, found in the
search, are shorter than the Nyiquist limit, it is probable that some of
them may be aliases of the true periods.

Many late type stars flare from time to time with typical duration of a
flare of less than one day. Such events should be eliminated before
period search, because points with large deviations from the mean value
have large weights in period searching algorithms. Unfortunately, the
ASAS observations are not well suited for identification of stellar
flares because they usually manifest themselves as single data points
indicating a substantial brightening. To get rid of them we rejected
data points deviating from the mean seasonal magnitude by more than 3.5
standard deviations, or from the grand mean value by more than 0.35~mag.

Apart from sudden brightenings many similar light drops occur within the
data. It is not clear what causes them. The simplest explanation relates
them to wrong measurements. Magnitude measurements fainter than the mean
value are more frequent than those above the mean (Pojma\'nski 2006,
private communication). We applied the same procedure to them as in case
of brightenings. Altogether, we rejected 229 drops and 87 brightenings,
which is about 0.6\% of all data points. Only one star, GJ 551 (=
Proxima) happened to brighten over two consecutive observations: first
by 1.13~mag above the mean level and 1.5 hour later it was still
brighter by 0.35~mag. In other cases we do not have neighboring
observations indicating significant brightness deviations from a normal
level. There were on average 285 measurements per star left, extended
over 6 seasons.

Late type active stars often show season-to-season light variations with
amplitudes of the order of 0.1~mag. A good example of such star is
AB~Dor which was observed nearly continuously over more than 20 seasons
(K\"urster \etal 1997, J\"arvinen \etal 2005). The long term variations
are usually interpreted as resulting from activity cycles. Several stars
from our sample also show seasonal variations. Because of that we have
done period search separately for each season unless the number of
useful observations in the particular season was lower than 50 data
points. Data from such seasons were merged with the ones from
neighboring seasons and analyzed together. A similar period search was
also performed on the whole data set after eliminating season to season
variations. We used the {\sc AoV} algorithm, together with the
subroutine {\sc caov} which can be  downloaded from the Alex
Schwarzenberg-Czerny homepage {\it http://www.camk.edu.pl/$\sim$alex/}.
A detailed description of the program can be found in
Schwarzenberg-Czerny (1989). In short, we are folding and binning data
points with a trial period, and for a given number of bins and data
points we obtain {\sc AoV} statistics 
$$\Theta_{\rm AoV}=s_1^2/s_2^2\eqno(1)$$
where
$$(r-1)s_1^2 =\sum\limits_{i=1}^{r}n_i(\overline{x_i}-\overline{x})^2\eqno(2)$$
and
$$(n-r)s_2^2=\sum\limits_{i=1}^{r}\sum\limits_{j=1}^{n_i}
(x_{ij}-\overline{x_i})^2\eqno(3)$$
with $r$ being a number of bins, $n_i$ -- number of data points in the
``$i$-th'' bin, $\overline{x_i}$ -- mean value of data points in the
``$i$-th'' bin and $\overline{x}$ -- mean value of all data points. We
used six phase bins and assumed that the detected periodicity is
significant when $\Theta_{\rm AoV}(P)\ge8.0$, unless the indicated period
was very close to 1 day due to a slow, systematic drift of the stellar
brightness over the whole season. Such periods were treated as spurious.
The results of the period search are presented and discussed in Section~3. 

\Section{Results and Discussion}
\subsection{Period Search}
Out of 180 analyzed stars only 31 showed statistically significant periodic
variations. Table~1 lists basic data for them and the results of the
period search. Spectral types, absolute magnitudes and X-ray data are taken
from NEXXUS. For X-ray variable stars average values of $\log R_x$ are
taken. Bolometric corrections were calculated from color indices $R-I$,
using a formula from Delfosse
\etal (1998): $BC=-0.083-0.501(R-I)-0.646(R-I)$.

\vspace*{-14pt}
\renewcommand{\arraystretch}{1}
\renewcommand{\TableFont}{\scriptsize}
\MakeTable{|@{\hspace{3pt}}l@{\hspace{3pt}}|@{\hspace{3pt}}c|
r@{\hspace{0pt}}c@{\hspace{0pt}}l|
r@{\hspace{0pt}}c@{\hspace{0pt}}l@{\hspace{3pt}}|
c@{\hspace{3pt}}|@{\hspace{3pt}}c@{\hspace{3pt}}|
@{\hspace{2pt}}c@{\hspace{2pt}}|@{\hspace{2pt}}r@{\hspace{0pt}}c@{\hspace{0pt}}l@{\hspace{2pt}}|
c|c|c|}{12.5cm}{Basic data for stars with
significant periodic variability}
{\hline
\uprule
name & ASAS & \multicolumn{3}{c|}{d} & \multicolumn{3}{c|}{$M_V$} & Sp & 
($\log L_b$) & ($\log R_x$)& \multicolumn{3}{c|}{P}    & $\Theta_{\rm AoV}$ &
amp  & mass\\
\dorule    & des. & \multicolumn{3}{c|}{pc} & \multicolumn{3}{c|}{mag}   &    
&            &           & \multicolumn{3}{c|}{days} & &mmag & \MS \\ 
\hline
\uprule
GJ 84      & 020504-1736.9 &  9&.&4 & 10&.&32 & M2.5 & 32.08 & $-4.4$ &44&.&51  & 24.4 &  26  & 0.45 \\
GJ 103     & 023422-4347.8 & 11&.&5 &  8&.&55 & M0   & 32.53 & $-2.8$ & 1&.&563 & 64.7 & 63 & 0.65 \\ 
GJ 1054A   & 030755-2813.2 & 18&.&0 &  8&.&96 & M0   & 32.36 & $-3.0$ & 0&.&513 & 15.4 & 18 & 0.63 \\ 
HIP 17695* & 034723-0158.3 & 16&.&3 & 10&.&53 & M3   & 31.98 & $-3.0$ & 3&.&880 & 25.1 &  39 & 0.43 \\ 
Gl 176     & 044256+1857.5 &  9&.&4 & 10&.&11 & M2   & 32.10 & $-4.7$ &38&.&92  & 19.5 & 34 & 0.48 \\ 
GJ 2036A   & 045331-5551.6 & 11&.&2 & 10&.&89 & M2   & 31.95 & $-3.0$ & 0&.&849 &  9.4 &   11 & 0.38 \\ 
GJ 182     & 045934+0147.0 & 26&.&7 &  7&.&96 & M0.5 & 32.75 & $-3.1$ & 4&.&410 & 19.9 &   27 & 0.69 \\ 
GJ 3331A   & 050650-2135.1 & 11&.&8 &  9&.&93 & M1.5 & 32.10 & $-3.0$ & 0&.&341 &   8.4 &   9 & 0.50 \\ 
GJ 205     & 053127-0340.6 &  5&.&7 &  9&.&18 & M1.5 & 32.37 & $-4.7$ &33&.&61  & 10.3 &   8 & 0.60 \\ 
GJ 3367    & 054717-0000.8 & 24&.&4 &  9&.&05 & M0   & 32.31 & $-3.4$ &12&.&05  &  8.9 &   14 & 0.62 \\ 
GJ 358     & 093946-4104.1 &  9&.&5 & 10&.&86 & M2   & 31.94 & $-3.9$ &25&.&26  & 21.3 &   14 & 0.39 \\ 
GJ 375     & 095834-4625.5 & 15&.&9 & 10&.&26 & M3.5 & 32.30 & $-3.1$ & 1&.&877 & 11.0 &   22& 0.37 \\ 
GJ 382     & 101217-0344.7 &  7&.&8 &  9&.&81 & M1.5 & 32.18 & $-4.7$ &21&.&56  & 21.6 &   11 & 0.52 \\ 
GJ 431     & 113146-4102.8 & 10&.&5 & 11&.&42 & M3.5 & 31.81 & $-3.5$ &14&.&31  & 10.5 &   31& 0.33 \\ 
TWA 5A     & 113156-3436.5 & 50& &  &  7&.&88 & M1.5 & 33.04 & $-3.1$ & 0&.&7767& 21.2  &   36& 0.70 \\ 
GJ 494     & 130046+1222.5 & 11&.&4 &  9&.&46 & M0.5 & 32.25 & $-3.4$ & 2&.&889 & 15.4 &   16  & 0.56 \\ 
GJ 551     & 142942-6240.8 &  1&.&3 & 15&.&49 & M5.5 & 30.76 & $-3.8$ &82&.&53  & 24.0 &   21 & 0.11 \\ 
GJ 569A    & 145429+1606.1 &  9&.&8 & 10&.&24 & M3   & 32.05 & $-3.7$ &13&.&68  &  8.3 &  10& 0.46 \\ 
GJ 9520    & 152152+2058.7 & 11&.&4 &  9&.&83 & M1.5 & 32.19 & $-3.3$ & 0&.&369 &  8.4 &   13& 0.51 \\ 
GJ 618A    & 162003-3731.7 &  8&.&5 & 10&.&96 & M3   & 31.89 & $-4.7$ &56&.&52  &  9.5 &   11 & 0.38 \\ 
GJ 2123A*  & 165648-3905.7 & 14&.&6 & 10&.&36 & M3   & 32.03 & $-2.5$ & 0&.&320 & 65.9 &   46 & 0.45 \\ 
GJ 669A    & 171954+2630.1 & 10&.&7 & 11&.&28 & M3.5 & 31.87 & $-3.2$ & 0&.&950 & 10.3 &   30 & 0.35 \\ 
GJ 674     & 172839-4653.7 &  4&.&5 & 11&.&10 & M3   & 31.80 & $-4.2$ &33&.&29  &  9.5 &   8 & 0.37 \\ 
GJ 729     & 184949-2350.2 &  3&.&0 & 13&.&09 & M3.5 & 31.21 & $-3.5$ & 2&.&869 & 15.3 &   10& 0.20 \\ 
GJ 799A/B  & 204151-3226.1 & 10&.&2 & 10&.&94 & M4.5 & 32.17 & $-3.0$ & 0&.&7813& 16.8 &   13 & 0.38 \\ 
GJ 803     & 204509-3120.5 &  9&.&9 &  8&.&82 & M1   & 32.52 & $-2.9$ & 4&.&848 & 30.5  &   46 & 0.64 \\ 
GJ 1264A   & 214905-7206.1 & 16&.&1 &  8&.&76 & M0.5 & 32.53 & $-3.3$ & 6&.&669 & 68.5 &   21 & 0.64 \\ 
Gl 841A    & 215741-5100.4 & 16&.&2 &  9&.&31 & M2.5 & 32.50 & $-3.2$ & 1&.&124 & 12.4 &   10 & 0.58 \\ 
GJ 867A    & 223845-2037.3 &  8&.&6 &  9&.&42 & M1.5 & 32.36 & $-3.2$ & 4&.&233 & 11.1 &   10 & 0.57 \\ 
GJ 890     & 230819-1524.6 & 21&.&8 &  9&.&18 & M0   & 32.24 & $-3.0$ & 0&.&431 & 11.0 &   37 & 0.60 \\ 
GJ 897A    & 233247-1645.4 & 15&.&4 & 10&.&01 & M2   & 32.22 & $-3.1$ & 4&.&828 & 12.0 &   14 & 0.49 \\ 
\hline									 
GJ 411     & 110321+3558.2 &  2&.&54& 10&.&46 & M2   & 31.89 & $-4.9$ &48&.&0   & - & -  &  0.44 \\
GJ 699     & 175749+0441.6 &  1&.&8 & 13&.&25 & M4   & 31.17 & $-5.4$ &130&.& &  - & - & 0.20 \\
\hline
\noalign{\vskip4pt}
\multicolumn{17}{p{11.5cm}}{Consecutive columns give names, ASAS
designations (= coordinates given in the order: RA-Dec.), distances,
absolute visual magnitudes, spectral types, bolometric luminosities,
$\log R_x=\log(L_x/L_{\rm bol})$, detected rotation periods, values of
the parameter $\Theta_{\rm AoV}$ at the listed period, amplitudes of
light variations, and estimated stellar masses.}
}

For two stars with no $R-I$ indices marked with asterisk the bolometric
corrections were calculated from the absolute visual magnitudes, as
suggested by Pettersen (1983): $BC=-0.397M_V+2.386$.

Stellar  masses were calculated from the absolute magnitude $M_V$ using
the formula derived by Delfosse \etal (2000): $\log(M/\MS)=
10^{-3}(0.3+1.87M_V+7.614M_V^2-1.698M_V^3+0.060958M_V^4).$

In case of tight binaries, the magnitudes given in Table~1 were not
corrected for fainter components, so their masses may be overestimated.
The exception is GJ 375 which consists of nearly identical components
(Montes \etal 2006). The formula given by Delfosse \etal (2000) applies
only to stars with $M_V>9$~mag. For a few brighter stars the masses were
found using the calibration given by Andersen (1991). The last two
entries in Table~1 are taken from literature.

Left panels of Figs.~1--4 present all photometric measurements of periodic
stars \vs ${\rm hjd=HJD-2\,450\,000}$~d with observations showing periodic
variations marked as filled symbols. Right panels show filled symbols
folded with the detected periods after applying corrections for seasonal
variations. In most cases periodic variability is seen over the whole
observational interval without any significant phase shift from one season
to another, but a few stars show transient periodic variability. The
number of filled symbols and the total number of analyzed observations are
given for each star.
 
Periods longer than 10 days were found only for 11 stars, \ie about 30\%,
of all detected periodic variables. We believe that such a low proportion
of slowly rotating stars is a result of observational selection: the
expected amplitudes of periodic variations of high activity stars are
substantially higher, hence easier to detect than in case of inactive
stars. Several more M type stars with short rotation periods of the order
of one or a few days are known in the literature (Messina \etal 2003) but
we found only two stars with long rotation periods: GJ 411 with $P_{\rm
rot}=48$~d (Noyes \etal 1984), and GJ 699 (Barnard star) with $P_{\rm
rot}\approx130$~d (Benedict \etal 1998). We include both stars into the
further analysis.

\begin{figure}[p]
\vglue-7mm
\hglue-7mm{\includegraphics[bb=100 200 500 700, width=14cm]{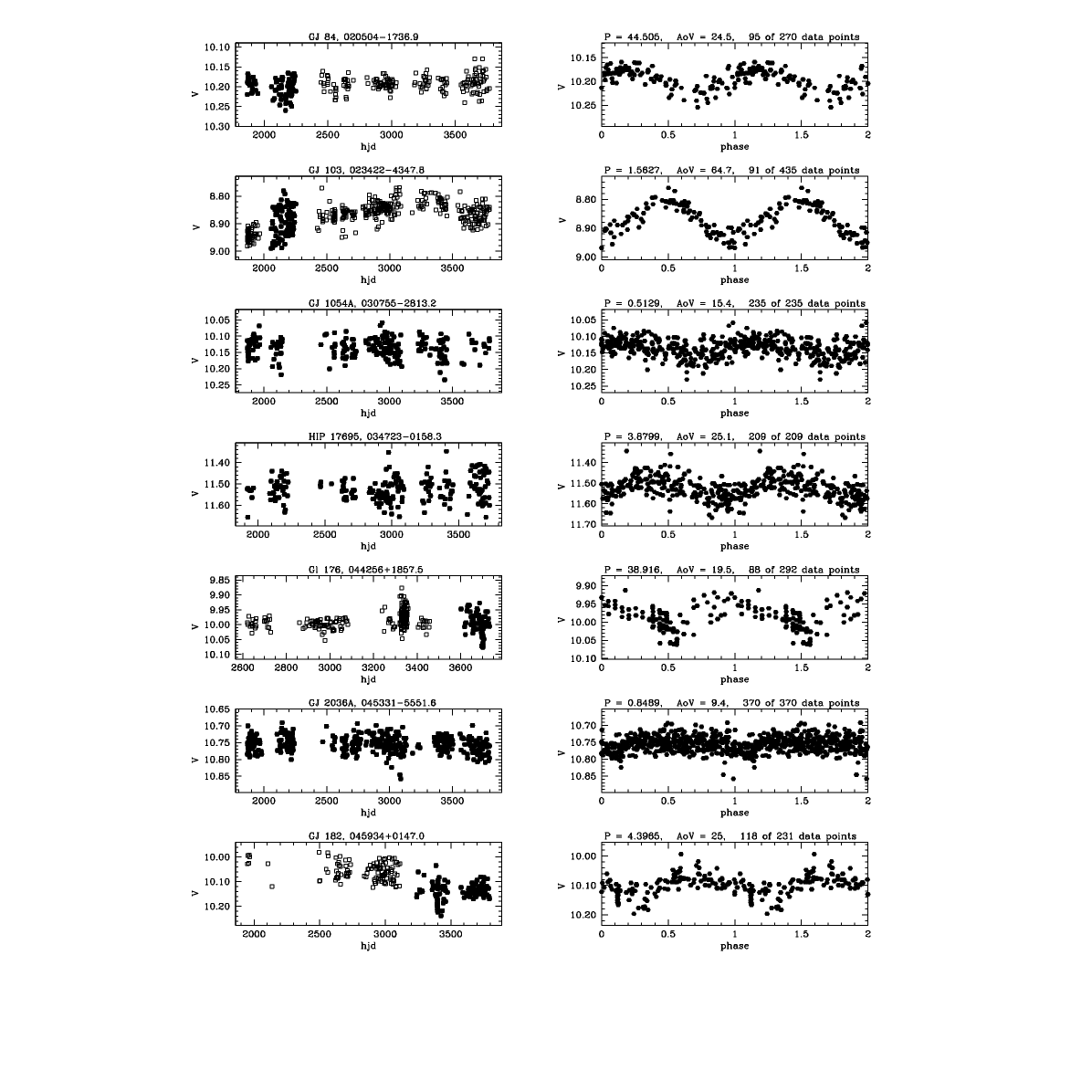}}
\vskip3mm
\FigCap{Photometric measurements of periodic stars \vs hjd = HJD
-2\,450\,000~d ({\it left panels}). Observations showing periodic variations
are plotted as filled symbols. {\it Right panels} show filled symbols
folded with the listed period. The brightness at right panels is
corrected for seasonal variations of luminosity. The number of filled
symbols and the total number of analyzed observations are also given.}
\end{figure}
\begin{figure}[p]
\vglue-1mm
\hglue-7mm{\includegraphics[bb=100 150 500 700, width=14cm]{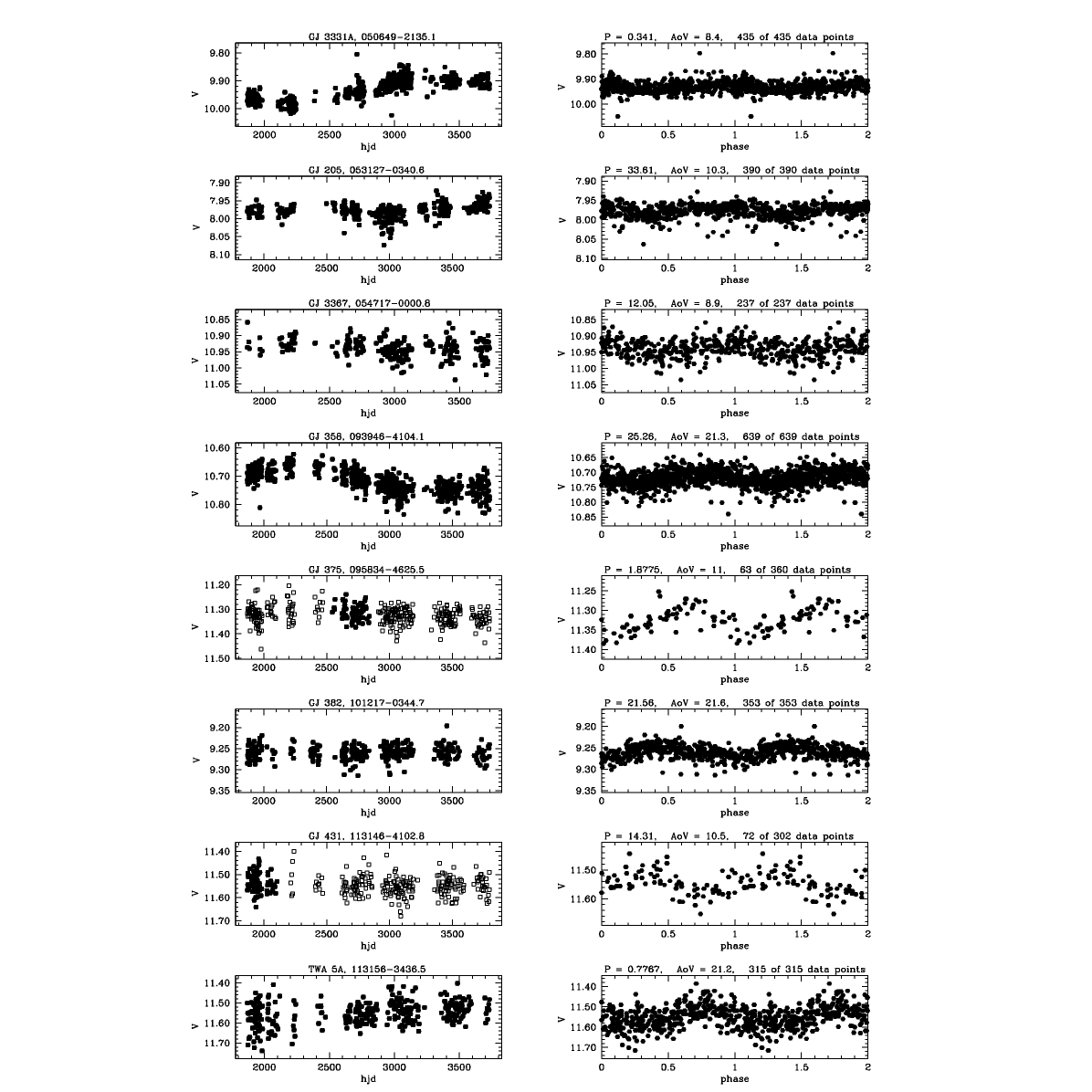}}
\vskip3mm
\FigCap{Same as Fig.~1}
\end{figure}
\begin{figure}[p]
\vglue-1mm
\hglue-7mm{\includegraphics[bb=100 150 500 700, width=14cm]{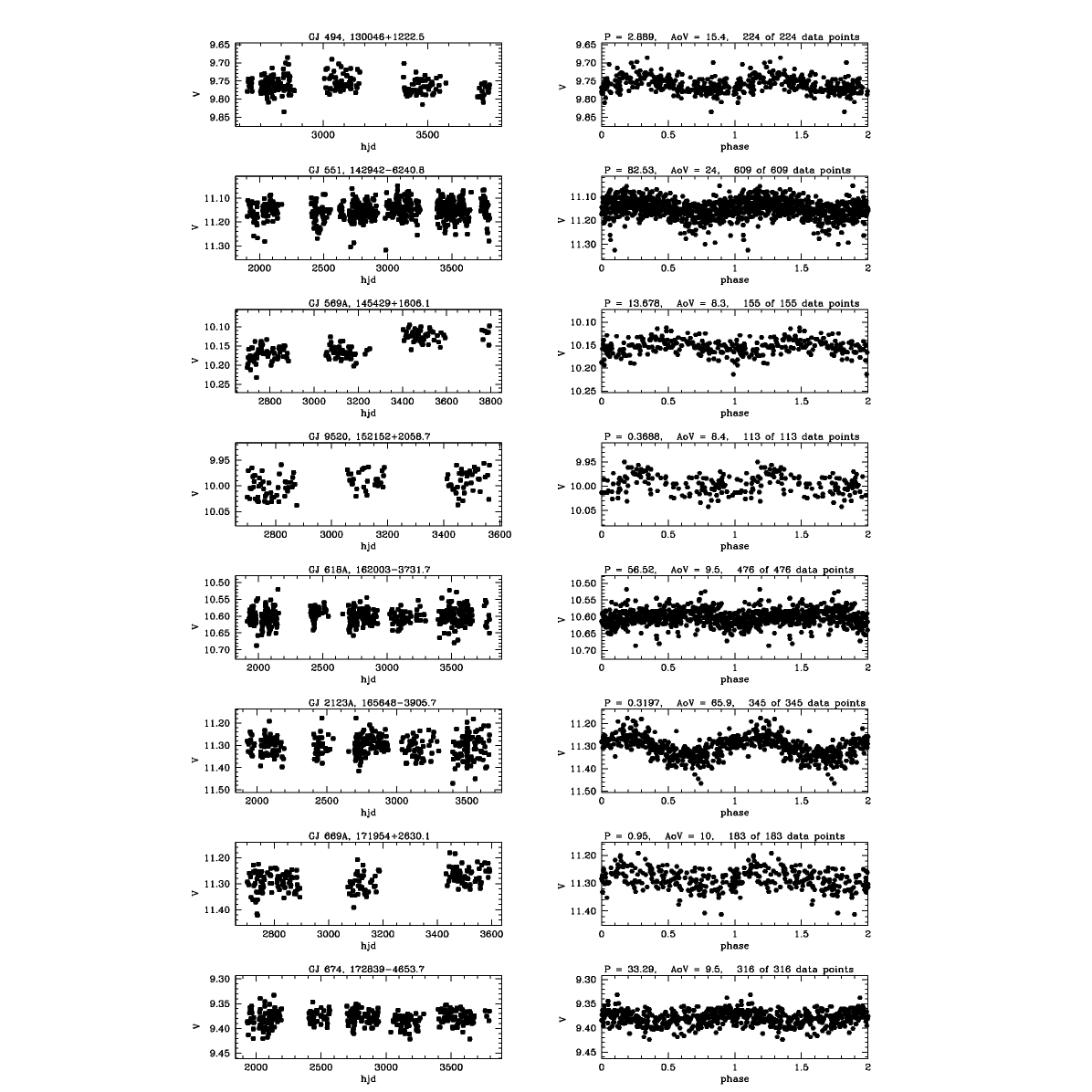}}
\vskip3mm
\FigCap{Same as Fig.~1}
\end{figure}
\begin{figure}[p]
\vglue-1mm
\hglue-7mm{\includegraphics[bb=100 150 500 700, width=14cm]{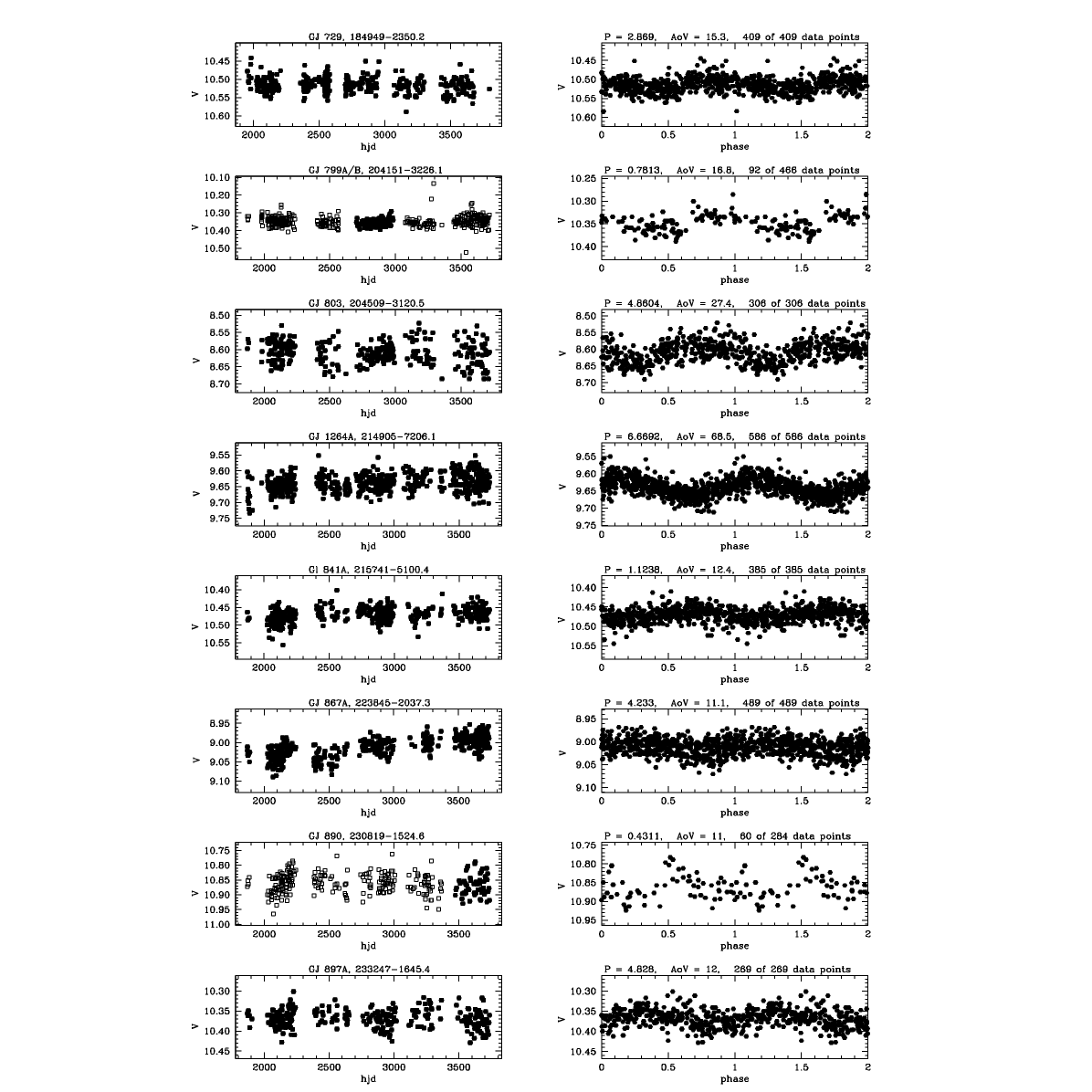}}
\vskip3mm
\FigCap{Same as Fig.~1}
\end{figure}

Notes on individual stars.

GJ~84 is a visual binary with the secondary component lying
0\zdot\arcs44 from the primary (Golimowski \etal 2004). The luminosity
difference is 4.6~mag in the F110W filter at the NICMOS camera, so the
contribution of the secondary is negligible, both in the {\it V}-band
and (most likely) in X-rays. Aliases 0.9749~d and 1.020~d are rather
excluded due to the low X-ray activity of the star.  

GJ~103 (CC~Eri) is a spectroscopic binary with components of spectral 
types K7 + M3 and the combined type M0 (NEXXUS). The photometric period
is the same as the orbital period (Evans 1959). 

GJ~1054A is a component of double line binary with an unknown period
(Gizis, Reid and Hawley 2002) which, however, according to the authors,
should be short. The most significant period in the ASAS photometry is
given in Table~1 (0.513~d, $\Theta_{\rm AoV}=15.4$), its alias 18.86~d is
also significant, ($\Theta_{\rm AoV}=13.3$) but we consider it as less
likely due to very high activity of the star.

GJ~182 is a single flare star with a rotation period known from literature
of 4.565~d (Byrne \etal 1984). We have not found any significant
variability at this period. Maximum value of $\Theta_{\rm AoV}=26.4$ is at
a period of 1.2921~d for the 106 data points between hjd 2496.9 and 3110.5,
however we consider the period of 4.410~d with $\Theta_{\rm AoV}=19.9$,
obtained for the whole data set, as the most probable.

GJ~2036A was observed together with GJ~2036B by ASAS. Component B is at
angular distance 8\arcs and is fainter by about 1~mag in the {\it
V}-band. The total $V$ magnitude of the binary as measured by ASAS is
10.75~mag. We attribute the photometric variability to the brighter
component.

GJ~3331A is the brightest component of the triple system. The pair BC is
8\zdot\arcs2 from the primary and the differences in magnitudes between
components A and BC are $\Delta V=0.722$, $\Delta R=0.574$, $\Delta
I=0.220$ (Jao \etal 2003). We assume that photometric variability is
related to GJ~3331A. Its parallax measurement has a large error (NEXXUS
database) so we do not include this star in Table~3.

GJ~205 -- aliases at 0.971~d and 1.028~d are rather excluded due to  low
activity of the star.

GJ~375 is a double line spectroscopic binary with a period of 1.875~d
and nearly identical components (Montes \etal 2006). Photometric period
is identical with the spectroscopic one. Mass, given in Table~1, refers
to one component.

TWA~5 -- appears to be a multiple system. This is a young system
belonging to the TW~Hya association. The main component consists of a
spectroscopic binary and the third component is separated by
0\zdot\arcs054 (Brandeker \etal 2003). All three stars are of spectral
type M1.5. In addition, there is also a brown dwarf at angular
distance of 2\arcs. Based on XMM observations  Argiroffi \etal (2005)
give $\log(L_X)= 29.8$ and $\log(L_X/L_{\rm bol})=-3.1$. for the triple M
dwarf system. The photometric period found in the ASAS data may probably
be related to the orbital period of a spectroscopic binary. Another
possible period is an alias 3.512~d. The mass given in Table~1 is based
on the absolute magnitude of the total system and, therefore, very
likely overestimated.

GJ~494 (DT Vir) has a low luminosity companion at angular separation
of 0\zdot\arcs48 (February 2000, Beuzit \etal 2004), fainter in the $K$
band by 4.4~mag, which makes its contribution completely negligible,
both in $V$ and in X-rays.

GJ~551 = Proxima has the lowest luminosity in our sample of  stars with
detected periods. ASAS photometric data indicate a clear periodicity of
82.5 days, very close to the rotation period found by  Benedict \etal
(1998). We have not found any significant signal at a period of 42 days,
visible in part of the data analyzed by Benedict \etal (1998), or around
30 days, reported by Guinan and Morgan (1996).

GJ~618A has a nearby companion (5\zdot\arcs4) of spectral type M5, which
is fainter in the {\it V}-band by 3.5~mag. This makes its photometric
contribution completely negligible and we can safely assume that the
photometric variability is related to the brighter component.
Nevertheless, the X-ray luminosity of M5 type star may contribute
significantly to the total X-ray emission of the binary if the star is
very active, so the value of $R_x=4.7$, listed in Table~1, should be
treated as an upper limit for the X-ray flux of GJ 618A.

GJ~2123A is a very active M3 star with an M4 companion about 1.8~mag
fainter in the {\it V}-band, lying at angular distance of 2\zdot\arcs8. We
attribute the photometric variability to the brighter component.

GJ~669A has a companion at angular distance of 16\arcs\!, fainter by about
1.6~mag in the {\it V}-band. Both stars are unresolved in the ROSAT and
ASAS data.

GJ~674, there are also significant peaks of AoV statistics at aliased
periods of 0.9709~d and 1.028~d with $\Theta_{\rm AoV}\approx10$. We prefer
33.29~d due to low activity of the star.

GJ~799 (AT Mic) is a visual binary with angular separation between
components of 3\zdot\arcs3, too close to be resolved either by ROSAT or by
ASAS. Both components have almost the same luminosity in {\it V}-band and
we assume that their X-ray luminosity is also the same. Values listed in
Table~1 are related to a single component.

GJ~803 (AU Mic) is a young, single star with rotation period of 4.865~d
(Torres \etal 1972). We see the strongest signal at a period of 4.8482~d
in the ASAS data (62 data points between  hjd 3450.9 and 3718.5), but
there is also a significant power at 4.8604 and 4.8617 d for the whole
data set. This may be related to the differential rotation of the star.

GJ~1264A has a companion at a distance of 1\zdot\arcs3, fainter by
1.2~mag in the {\it V}-band. The stars are unresolved in ROSAT and ASAS
data. We assume that all activity and variability is related to the
brighter component.

GJ~841A is a spectroscopic binary with a period of 1.1248~d  (Jeffries
and Bromage 1993). A similar period of 1.1237~d is visible in the whole
ASAS data set (with the {\sc AoV} statistics equal to 12.4). More
significant (aliased) period of 0.5283~d with $\Theta_{\rm AoV}=12.5$ 
does not fit to the  Jeffries and Bromage data.

GJ~867A (FK Aqr) has a companion separated by 23\zdot\arcs5 and fainter
by 1.9~mag in the {\it V}-band. The stars are unresolved in ROSAT and
ASAS data. GJ~867A is a spectroscopic binary with an orbital period of
4.083~d (Herbig and Moorhead 1965). Bopp and Espenak (1977) also found a
photometric period of 4.08~d, but Byrne \etal (1987) claim a photometric
period of 4.39~d. Cited above photometric periods are  close to, but not
identical with our value 4.232~d. The mass given in Table~1 is likely
overestimated.

GJ~890 (HK Aqr). Barnes and Collier Cameron (2001) give a rotation
period of 0.4307~d for this star. We obtained a similar period of 0.4311~d.

GJ~897A has a companion separated by 0\zdot\arcs5 and fainter by 0.5~mag
in the {\it V}-band. The stars are not resolved by ROSAT and ASAS. We
attribute the photometric variability to the brighter component.

\subsection{Empirical Turnover Times}
To derive the empirical turnover times for low mass stars the procedure
applied by Stępie\'n (1994, 2003) was used, except that the analyzed
stars were grouped in the present paper according to mass rather than
$B-V$. Natural breaks in the mass distribution were used for this
grouping. Two stars: GJ~182 and TWA~5A stand out as the most massive
stars. Because no slowly rotating stars with similar masses occur in
Table~1, the stars were excluded from the further analysis. 10 ``high
mass'' stars have masses between 0.56~\MS\ and 0.65~\MS, 10 ``medium
mass'' stars have masses between 0.43~\MS\ and 0.52~\MS\ and 8 ``low
mass'' stars have masses between 0.33~\MS\ and 0.39~\MS. Two stars with
masses 0.2~\MS\ and one with 0.11~\MS\ are termed ``very low mass'' stars
(Table~2).
\renewcommand{\arraystretch}{1}
\renewcommand{\TableFont}{\footnotesize}
\MakeTable{|l|c|c|c|c|c|}{12.5cm} {Values of the coefficients $a$ and
$b$ from Eq.~(4) for four mass intervals and the resulting turnover times}
{\hline
stars         & av. mass & No. of stars & $a$  & $b$  & $\tau_c$(d)\\
\hline
high mass     & 0.61 & 10 & $-2.94\pm0.08$ & $-0.050\pm0.006$ & $30\pm4$\\
medium mass   & 0.47 & 10 & $-3.02\pm0.18$ & $-0.040\pm0.007$ & $38\pm6$\\
low mass      & 0.37 & 8  & $-3.14\pm0.08$ & $-0.029\pm0.003$ & $53\pm6$\\
very low mass & 0.17 & 3  & $-3.00$        & $-0.016\pm0.005$ & $95\pm30$\\
\hline}

\begin{figure}[htb]
\centerline{\includegraphics[width=10cm]{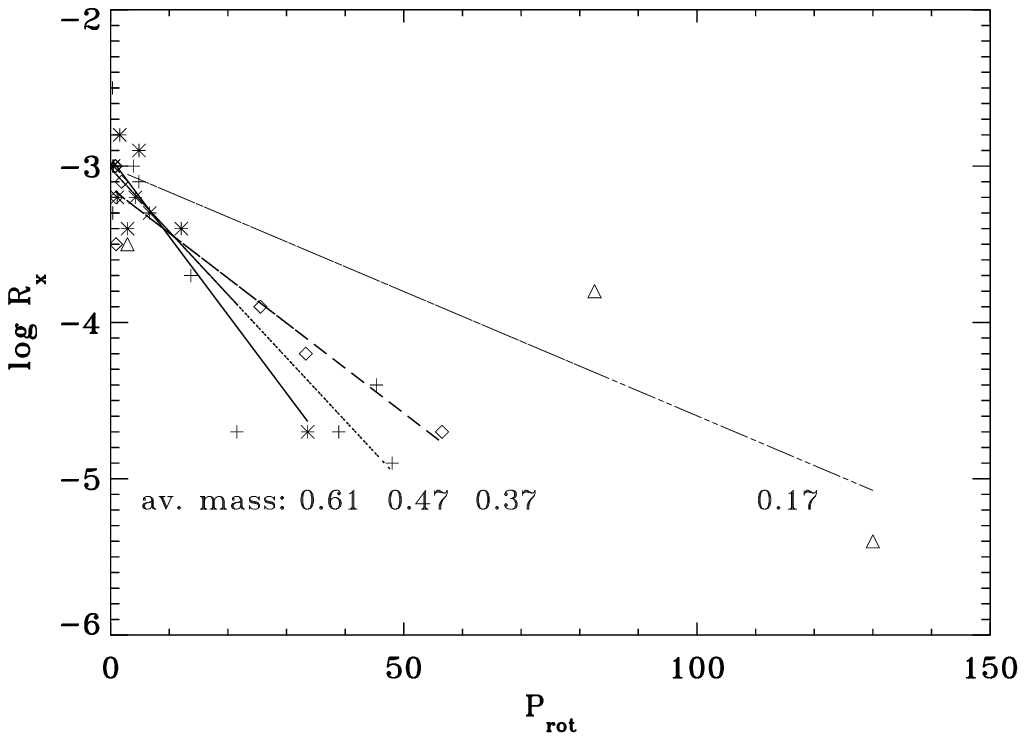}}
\FigCap{X-ray to bolometric flux ratio \vs rotation period for stars
from Table~1. Straight lines are the least square fits to observational data
grouped according to mass. Asterisks correspond to high mass stars,
pluses to medium mass, diamonds to low mass and triangles to very low
mass stars. Average mass for each group is indicated at the fitted
line.}
\end{figure}
Fig.~5 presents values of $\log R_x$ \vs $P_{\rm rot}$ for all four
groups of stars. Different groups are marked with different symbols.
Linear relations of the form 
$$\log R_x=a+bP_{\rm rot}\eqno(4)$$
are fitted to the data within each group. For the first three groups
both coefficients were determined with the least squares method. Because
only one short period star with very low mass is plotted in Figs.~1--4 the
value of $a$ in this case was assumed to be equal to $-3$. Table~2
gives values of the coefficients from Eq.~(1) for each group, together
with their errors. The value of the coefficient $b$ for very low
mass stars is extremely uncertain due to a low number of data.

Values of empirical turnover time can be found from relation: $\tau_c=
-cb^{-1}$, where the normalization factor $c$ remains undetermined
unless one wants to compare the obtained values of $\tau_c$ to values
obtained with another method (for details, see Stępie\'n 1994). A value
of the normalization coefficient $c=1.515$ was found by fitting the
presently calculated turnover times, supplemented with the values
obtained by Stępie\'n (1994) from the X-ray data of G--K type stars
($B-V$ between 0.55~mag and 1.25~mag), to the relation determined by
Stępie\'n (2003) from CaII emission. The latter relation was found by
using correctly reduced values of the net calcium emission flux,
contrary to, so called, excess calcium emission flux used earlier by
Stępie\'n (1994). The last column of Table~2 gives normalized values of
$\tau_c$ and Fig.~6 shows $\tau_c$ \vs mass. The relation from
Stępie\'n (2003) is plotted as a solid line and the values of turnover
times found from X-ray data are given with their errors. The older data
are plotted with $B-V$ values transformed to stellar mass. It is seen
from Fig.~6 that, after passing a plateau for masses in the range
0.8~\MS--0.6~\MS, turnover time increases again for still lower masses.
The increase is not as steep as resulting from the extrapolation of the
theoretical values (Kim and Demarque 1996). Overplotted are values of
turnover times found by Pizzolato \etal (2003). They used a different
sample of stars and a somewhat different method but the agreement
between both sets of data is satisfactory. They were not able to
determine turnover times for stars less massive than 0.6~\MS\ due to a
shortage of data for slowly rotating stars.
\begin{figure}[htb]
\centerline{\includegraphics[width=9.6cm]{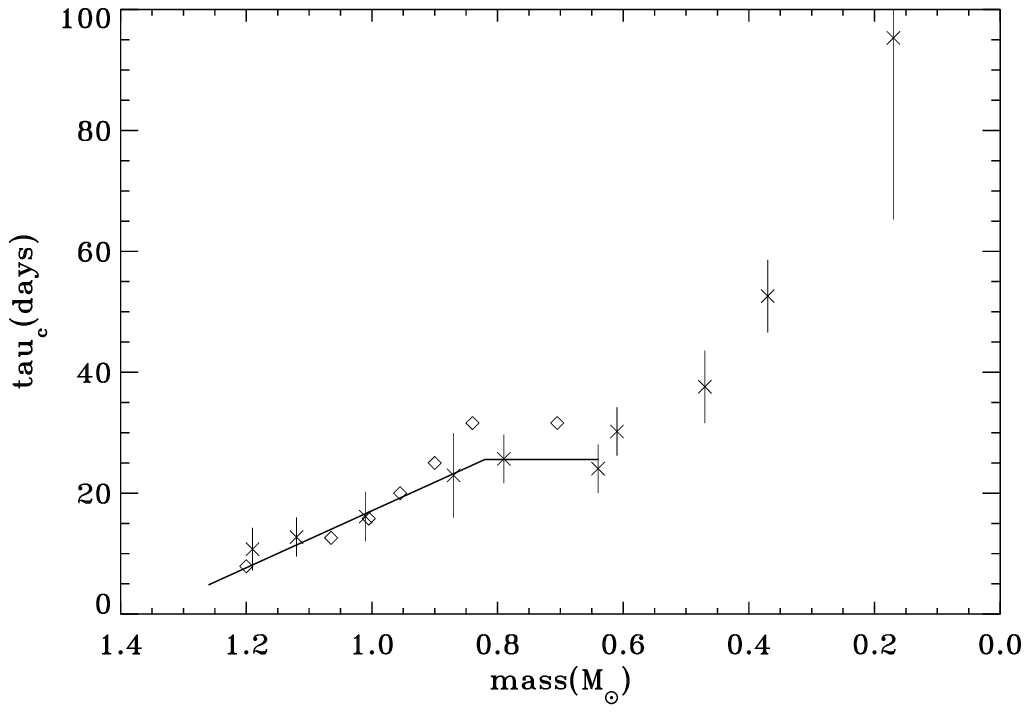}}
\FigCap{Empirical turnover times with error bars, determined from X-ray
fluxes, \vs stellar mass. Data from Table~2 are supplemented with older
determinations for more massive stars (Stępie\'n 1994). Values of the
turnover times are scaled to data found from calcium emission fluxes
(solid line, Stępie\'n 2003). Diamond give values of empirical turnover
times from Pizzolato \etal (2003).}
\end{figure}

\begin{figure}[htb]
\centerline{\includegraphics[width=10cm]{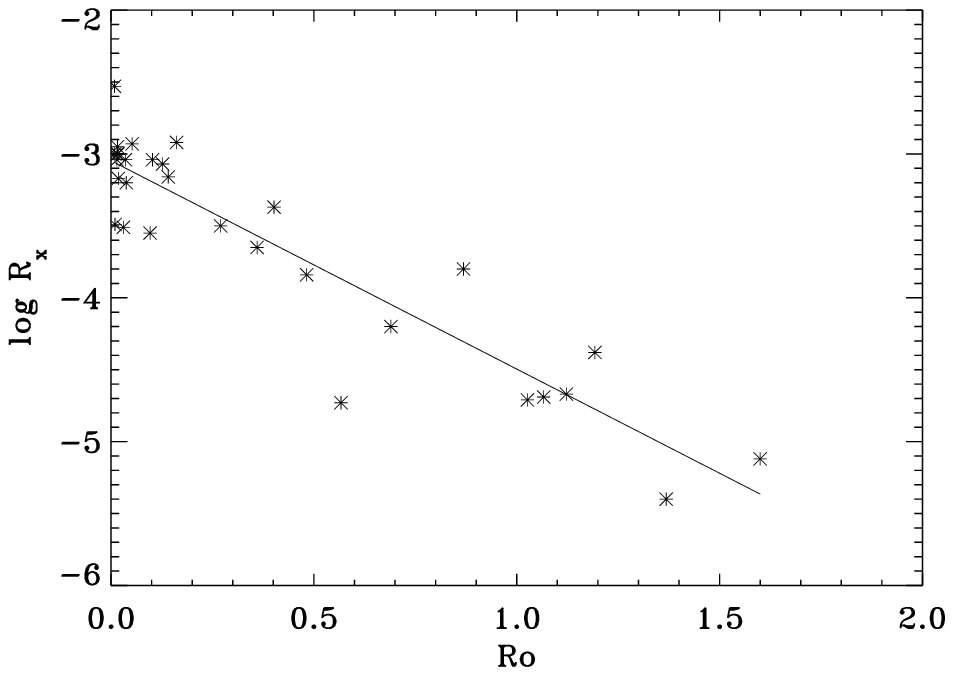}}
\vglue-3mm
\FigCap{X-ray to bolometric flux ratio \vs Rossby number $Ro= P_{\rm
rot}/\tau_c$ calculated at the base of empirical convective turnover
times from the Table~2. The straight line gives the best fit (see
text).}
\end{figure}
Fig.~7 shows the X-ray data of the stars from Fig.~5 plotted \vs Rossby
number which was calculated for each star using turnover time from
Table~2. As expected, the scatter is considerably lower than in Fig.~5
and the exponential relation between $R_x$ and $Ro$ seems to describe
satisfactorily the fit $\log(R_x)=-(3.05\pm0.07)-(1.45\pm 0.12)Ro$.
Similar relation has been shown to hold for G and K type stars
(Stępie\'n 1994: $\log(R_x)=-(3.71\pm0.27)-(1.46\pm 0.13)Ro$). The free
term differs somewhat between both relations but the slope agrees very
well.

\subsection{Age--Period and Period--Activity Relations}
\vspace*{9pt}
Active stars from the lower MS are expected to lose angular momentum {\it
via} a magnetized wind. Observations of stellar clusters show indeed that
single stars spin down with age. Unfortunately, accurate measurements of
rotation periods are available for members of young clusters only. In case
of M\,67, the oldest cluster (4~Gyr) for which useful measurements exist,
no more than just $\vv\sin i$ values have been measured for some stars. The
available data may be sufficient to calibrate the age--rotation relation
for solar type stars but M type stars evolve substantially slower and we
need accurate data for stars much older than members of the investigated
clusters. Such stars are expected to exist among field stars in the solar
vicinity but accurate determination of field star age is difficult. We are
forced to use proxies only statistically related to age. A commonly used
proxy of this kind is kinematic class (Eggen 1969) based on the ($U$, $V$)
plane, where $U$, $V$ and $W$ are the components of space motion of a
star. Table~3 gives star names, components of proper motions in right
ascension and declination, radial velocity, the calculated values of the
$U$, $V$ and $W$ components for stars from Table~1 for which complete data,
including radial velocity, are available. Proper motions and radial
velocities are from the NEXXUS database, {\it U}-axis is directed opposite
to the Galactic center, {\it V}-axis -- in direction of Galactic rotation,
{\it W}-axis -- in direction of the North Galactic Pole. YD means Young
Disk with kinematical parameters $-20<U<50$, $-30<V<10$, $-25<W<10$, OD --
Old Disc, YD/OD Young or Old Disc, OD/H -- Old Disk or Halo.
\MakeTable{|l|
r@{\hspace{0pt}}c@{\hspace{0pt}}l|
r@{\hspace{0pt}}c@{\hspace{0pt}}l|
r@{\hspace{0pt}}c@{\hspace{0pt}}l|
r@{\hspace{0pt}}c@{\hspace{0pt}}l|
r@{\hspace{0pt}}c@{\hspace{0pt}}l|
r@{\hspace{0pt}}c@{\hspace{0pt}}l|
c|}{12.5cm}{Kinematic properties of stars}
{\hline
\uprule name &
\multicolumn{3}{c|}{$\mu_{\alpha} $} &
\multicolumn{3}{c|}{$\mu_{\delta}$} & 
\multicolumn{3}{c|}{$V_{\rm rad} $} & 
\multicolumn{3}{c|}{$U$} & 
\multicolumn{3}{c|}{$V$} & 
\multicolumn{3}{c|}{$W$} & remarks \\
   & 
\multicolumn{3}{c|}{[mas]}  & 
\multicolumn{3}{c|}{[mas]}  & 
\multicolumn{3}{c|}{[km/s]} & 
\multicolumn{3}{c|}{[km/s]} & 
\multicolumn{3}{c|}{[km/s]} & 
\multicolumn{3}{c|}{[km/s]} & \dorule\\
\hline
\uprule
GJ 84     & $ 1317$&.&5 & $ -173$&.&9 & $ 23$&.&3 & $ 45$&.&2 & $-44$&.&5 & $ -6$&.&8 & OD \\  
GJ 176    & $  659$&.&8 & $-1114$&.&7 & $ 26$&.&0 & $ 22$&.&6 & $-57$&.&4 & $-14$&.&7 & OD \\
GJ 182    & $   37$&.&2 & $  -93$&.&9 & $ 18$&.&2 & $ 11$&.&3 & $-16$&.&8 & $ -9$&.&2 & YD \\
GJ 2036A  & $  132$&.&9 & $   73$&.&9 & $ 40$&.&6 & $  8$&.&4 & $-35$&.&1 & $-20$&.&3 & YD/OD\\
GJ 205    & $  763$&.&0 & $-2092$&.&8 & $  8$&.&5 & $-22$&.&2 & $-55$&.&6 & $-10$&.&2 & OD  \\
GJ 3331A  & $   49$&.&1 & $  -32$&.&7 & $ 20$&.&0 & $ 11$&.&5 & $-14$&.&1 & $ -9$&.&0 & YD \\
GJ 3367   & $  -85$&.&2 & $  -77$&.&3 & $-23$&.&0 & $-24$&.&4 & $  8$&.&0 & $ -6$&.&8 & YD \\
GJ 358    & $ -526$&.&1 & $  356$&.&9 & $  7$&.&0 & $ 28$&.&5 & $ -6$&.&7 & $ -2$&.&8 & YD \\
GJ 382    & $ -152$&.&9 & $ -242$&.&9 & $  7$&.&6 & $  2$&.&4 & $-12$&.&5 & $ -2$&.&8 & YD \\
GJ 431    & $ -715$&.&8 & $  170$&.&8 & $  6$&.&0 & $ 32$&.&8 & $-17$&.&2 & $ -1$&.&2 & YD \\
GJ 494    & $ -618$&.&8 & $  -16$&.&6 & $-11$&.&7 & $ 29$&.&3 & $-17$&.&2 & $-10$&.&3 & YD \\
GJ 551    & $-3775$&.&4 & $  769$&.&3 & $-21$&.&6 & $ 28$&.&9 & $  1$&.&5 & $ 13$&.&7 & YD \\
GJ 569A   & $  276$&.&1 & $ -122$&.&4 & $ -7$&.&2 & $ -7$&.&8 & $  3$&.&2 & $-13$&.&3 & YD \\
GJ 9520   & $   78$&.&4 & $  129$&.&4 & $  6$&.&5 & $ -0$&.&4 & $  9$&.&5 & $  4$&.&3 & YD \\
GJ 618A   & $ -729$&.&3 & $  991$&.&2 & $ 43$&.&0 & $-35$&.&6 & $ -3$&.&2 & $ 54$&.&9 & OD \\
GJ 2123A  & $   65$&.&8 & $ -109$&.&4 & $-18$&.&0 & $ 17$&.&8 & $  1$&.&3 & $ -9$&.&0 & YD \\
GJ 669 A  & $ -221$&.&2 & $  349$&.&0 & $-33$&.&6 & $ 34$&.&5 & $-19$&.&1 & $ -3$&.&6 & YD \\
GJ 674    & $  573$&.&4 & $ -879$&.&6 & $-10$&.&0 & $ 14$&.&5 & $ -5$&.&1 & $-19$&.&3 & YD \\
GJ 729    & $  637$&.&6 & $ -192$&.&5 & $-11$&.&5 & $ 13$&.&0 & $ -1$&.&2 & $ -7$&.&1 & YD \\
GJ 799AB  & $  269$&.&3 & $ -365$&.&7 & $  5$&.&0 & $  2$&.&7 & $-15$&.&5 & $-16$&.&2 & YD \\
GJ 803    & $  280$&.&4 & $ -360$&.&1 & $ -4$&.&6 & $ 10$&.&2 & $-16$&.&4 & $-10$&.&4 & YD \\
GJ 1264A  & $  300$&.&8 & $ -291$&.&1 & $ -4$&.&5 & $ 26$&.&0 & $-19$&.&2 & $ -1$&.&7 & YD \\
GJ 867A   & $  450$&.&6 & $  -79$&.&9 & $ -8$&.&7 & $ 17$&.&8 & $-10$&.&3 & $ -2$&.&1 & YD \\
GJ 890    & $  101$&.&6 & $  -24$&.&1 & $  7$&.&0 & $  6$&.&2 & $ -3$&.&1 & $-10$&.&9 & YD \\
GJ 897A   & $  317$&.&0 & $ -234$&.&0 & $ -2$&.&6 & $ 13$&.&1 & $-24$&.&6 & $ -7$&.&8 & YD \\
GJ 411    & $ -580$&.&5 & $-4770$&.&0 & $-84$&.&7 & $-46$&.&2 & $-53$&.&8 & $-74$&.&3 & OD \\
GJ 699    & $ -798$&.&7 & $10337$&.&8 &$-110$&.&6 & $141$&.&1 & $  4$&.&5 & $ 18$&.&2 & OD/H\\
\hline
\noalign{\vskip4pt}
\multicolumn{20}{p{10.5cm}}{The columns give name of a star, proper 
motion in right ascension and declination, radial velocity, three galactic
components of velocity, kinematical class (see text)}
}

Known spectroscopic binaries with short periods were excluded because
their activity levels and variability periods (assuming synchronization
of orbital and rotational period) are not related to ages. The last
column gives kinematic class.  We identify 20 stars belonging to young
disk (YD), 1 star falling at the border between young and old disk
(YD/OD) and 4 stars belonging to old disk (OD). Both stars added to our
sample from literature belong to old disk (Fleming \etal 1995), so we
have all in all 6 OD stars, although GJ 699 may even belong to halo
(Delfosse \etal 1998).

As it is seen from Table~3, all short period stars belong to YD or
YD/OD. However, there are also 4 YD stars in our sample with periods
longer than 10 days, including Proxima with a period of 82.5~d -- the
longest one found in our search. Such a mixture of rapidly and slowly
rotating stars is not surprising. YD stars are on average about 3--4~Gyr
old (Meusinger, Stecklum and Reimann 1991) but individual stars may have
ages up to several Gyr. If Proxima has a common origin with $\alpha$~Cen
(Wertheimer and Laughlin 2006), its age is about 6~Gyr, yet its
kinematical characteristics are typical of young disk.

The average period of YD stars is 9.4 d and the median value is about
2~d. Contrary to YD stars, we do not have any rapidly rotating stars
among OD objects. All six stars have periods longer than 30~d with a
mean value of 58.8~d and a median of 47~d. We assume that OD stars have age
of the order of 10~Gyr (Meusinger \etal 1991).

Comparison of both median values indicates that angular momentum of M
type stars decreases, on average, by a factor of 25 in 10 Gyr but this
factor may vary between 5 and 100 if the initial rotation periods of
individual stars are taken from observations of young clusters (Barnes
2003).

Based on observations of $\vv\sin i$ of solar type stars Skumanich
(1972) found an empirical relation describing the dependence of the
rotation period on time: $P_{\rm rot}\propto t^{1/2}$. More numerous
observations of stars in several young clusters showed that the
Skumanich law breaks down for rapidly rotating stars which show the
effect of saturation, visible as a flattening of $P_{\rm rot}(t)$ when
$t\rightarrow0$. Considering slowly rotating stars we still may ask, how
universal the Skumanich law is? Is it mass independent? A number of
formulas describing spin down rate of single stars have been suggested
in the literature. One of them, derived by Kawaler (1988), has a form:
${\rm d}\omega/{\rm d}t=A\omega^3$, where the coefficient $A$ depends 
on stellar mass and radius. Neglecting time variability of stellar
parameters we recover the Skumanich law: $\omega\propto At^{-1/2}$.
Here, the dependence on stellar parameters appears only through the
coefficient of proportionality $A$ whereas the functional dependence on
time ($t^{-1/2}$) is mass independent. Sills \etal (2000) assumed that
the Kawaler formula can be applied even to VLMS. Another spin down
formula, in which mass dependence occurs only {\it via} an exponential term
containing turnover time, was derived by Stępie\'n (1988, 2006)
$$-\frac{{\rm d}\omega}{{\rm d}t}=7\times 10^{-9}\omega{\rm
e}^{-Ro/0.335}\eqno(5)$$ 
where $\omega$ is expressed in 1/d, and $t$ in years. After integration the
above formula predicts different functional dependence of the rotation
period on time for different turnover times, hence masses. In particular,
stars with turnover times substantially longer than the Sun spin down
faster and rotate slower at the same age. Fig.~8 shows the time variability
of a rotation period for stars with turnover times from Table~2 and
(assumed) initial rotation period of 1~d. For comparison, $P_{\rm rot}(t)$
of a star with turnover time corresponding to the solar value (15.5~d) and
to the value from plateau (25.5~d) are also given.  Overplotted are the
observed rotation periods of the Sun, Proxima and all six OD stars (plotted
at an age of 9.9~Gyr for better visibility).  The observational point at an
age of 4 Gyr corresponds to a period of 17~d, calculated from the average
value of $\vv\sin i$ of four solar type stars with $0.60~{\rm
mag}<B-V<0.69~{\rm mag}$ measured by Pace and Pasquini (2004). Two highest
lying curves correspond to $\tau_c= 95$~d and to this turnover time minus
its estimated error, \ie $\tau_c=65$~d. It seems that the lower value
better agrees with observations of the least massive stars.  The overall
agreement with observational data supports the slow down model described by
Eq.~(5). Note that the rotation period of a star with the initial period of
a fraction of a day would vary along the curve lying correspondingly lower
in Fig.~8 than the curve starting at $P_{\rm init}=1$~d.
\begin{figure}[htb]
\centerline{\includegraphics[width=10.5cm]{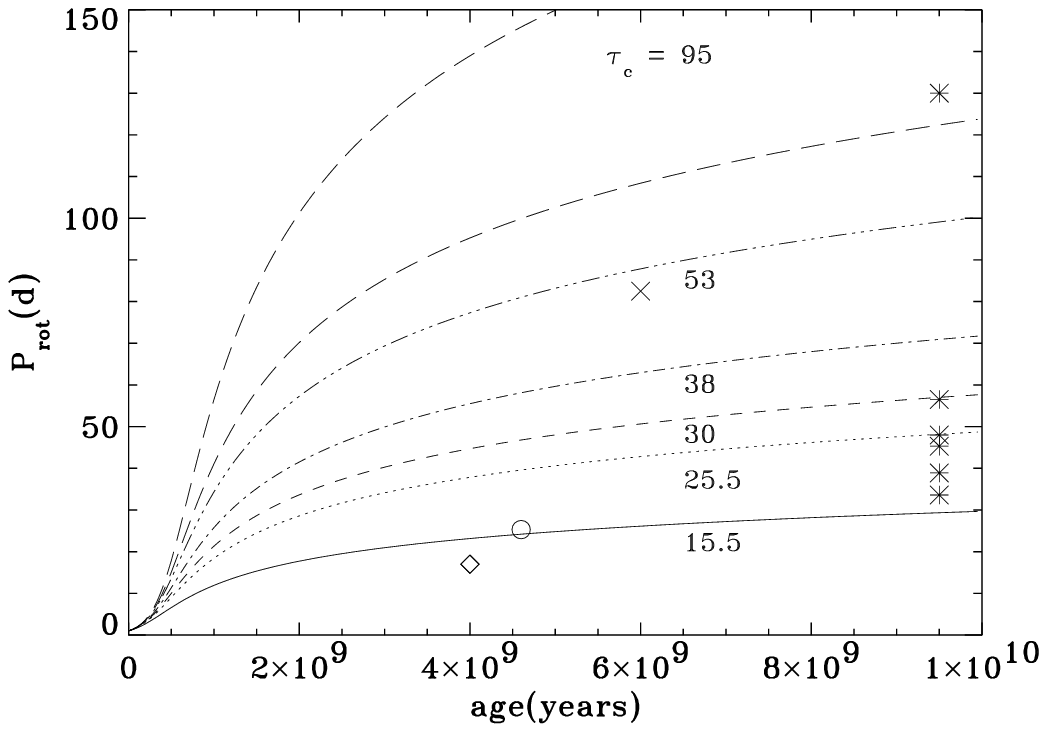}}
\vskip2mm
\FigCap{Time evolution of stellar rotation period, resulting from
integration of Eq.~(5) with the initial value of 1~d. Different curves
correspond to different turnover times, as indicated. Overclouded are
observational data with the following coding: diamond -- the average
rotation period of four solar type members of M\,67, calculated from their
$\vv\sin i$ values, open circle -- the Sun, cross -- Proxima and asterisks
-- six old disk stars from Table~3.}
\end{figure}

Extensive observations of chromospheric emission lines and X-ray fluxes
of solar type stars in many open clusters resulted in well determined
age--activity relations (\eg Barry 1988, Soderblom, Duncan and Johnson
1991, Haisch and Schmitt 1996, G\"udel, Guinan and Skinner 1997).
However, a direct extrapolation of a relation calibrated on stars with
masses 0.7--1.1~\MS\ to M stars with masses several times lower is not
obvious {\it a priori} and needs to be examined. Observations of M stars
of a given age show that rotation rate and activity level increase
towards later spectral subtypes (Delfosse \etal 1998, Scholz and
Eisl\"offel 2005) but a mean level of activity decreases with age for a
given spectral type (Hawley, Reid and Tourtellot 2000). Similarly as in
case of rotation data, the activity measurements of stars with well
determined ages are restricted to nearby open clusters and the
observations of activity level of still older M type stars with ages of
the order of 10~Gyr refer to field stars. Two such searches have
recently been performed: one in H$\alpha$ (Silvestri, Hawley and Oswalt
2005) and the other in X-rays (Feigelson \etal 2004). Silvestri \etal
(2005) observed 139 M stars in wide binaries with white dwarf
companions. Ages of the binaries were determined from the cooling age of
the white dwarf components. Most of the investigated M dwarfs have ages
below 4~Gyr but there are also many stars with ages up to 10~Gyr
(regretfully the authors do not list ages of individual stars). The
results confirm earlier findings that the mean activity level increases
with decreasing mass. The analysis of kinematics of the investigated
stars showed that active stars (of dMe type) have lower space velocities
and are confined to thin disk. Inactive stars (of dM type) have
systematically higher space velocities, characteristic of thick disk.
Feigelson \etal (2004) analyzed the results of a deep, pencil-beam X-ray
observations in a Galactic high-latitude field, obtained with Chandra.
The observations of late-type MS stars are compared with predictions
based on convolution of X-ray luminosity function with the known spatial
distribution of old disk stars. The authors came to the conclusion that
stellar X-ray luminosities decrease with age like $L_x\propto t^{-2}$
over $1<t<11$~Gyr, rather than $L_x\propto t^{-1}$ resulting from simple
power laws: $\vv_{\rm rot}\propto t^{-1/2}$ (Skumanich 1972) and $L_x
\propto\vv_{\rm rot}^2$ (Pallavicini \etal 1981, Pizzolato \etal 2003). 
The result shows that at least one the above power laws does not apply
to M stars which dominate stellar source counts in the search by
Feigelson \etal (2004).

Stępie\'n (1988, 1994) suggested an alternative period-activity
relation, in which the activity level depends exponentially on the
Rossby number.

We have shown in the previous section that the X-ray activity of M stars
 decreases with the Rossby number similarly to that of G--K type stars
\ie
$$\log R_x\propto-(1.45\pm 0.12)Ro.\eqno(6)$$

Quality of fit of both, power and exponential relations to observational
data is similar (Hempelmann \etal 1995) so it is not possible to
discriminate between them on the basis of the rotation-activity diagram
alone. However, the age--activity relation resulting from Eq.~(6),
together with Eq.~(4), is in a much better agreement with the results
obtained by Feigelson \etal (2004) than the power relation  $L_x\propto
t^{-1}$.
 
Using turnover times given in Table 2, Rossby numbers of six old disk
stars were calculated and the average value obtained: $\overline{Ro}=
1.2\pm0.4$. The uncertainty of this value comes from errors of turnover
times and the scatter of individual Rossby numbers. Assuming that the
obtained value of  $\overline{Ro}$ is typical for old M type stars and
that young M type stars have $Ro\ll1$, we obtain:
$$(\log R_x)_{\rm young}-(\log R_x)_{\rm old}=1.7\pm0.50.\eqno(7)$$

X-ray observations of Hyades (Reid \etal 1995) and Praesepe (Franciosini
\etal 2003), open clusters with ages about 600~Myr and 800~Myr
respectively, show their M stars still in saturation regime. Assuming the
age of M stars at the end of saturation regime $\approx1$~Gyr and of old
stars $\approx10$~Gyr, the 10-fold increase in age should result, according
to Eq.~(7), in a decrease of X-ray flux by a factor $10^{-1.7}\!\!$. This
prediction can directly be compared with observations.  All six OD stars
were observed in X-rays (see Table~1) and their average value of $\log R_x$
is equal to $-4.84$. The average value of $\log R_x$ of young, rapidly
rotating stars is about $-(3.0{-}3.2)$ (Pizzolato
\etal 2003, see also Fig.~5), so the observed difference $(\log
R_x)_{\rm young}-(\log R_x)_{\rm old}\approx1.6{-}1.8$ is in excellent
agreement with the predictions based on Eq.~(7) and in a fair agreement
with the results by Feigelson \etal (2004). 

Based on the results of the present paper we can explain why M type
stars show higher activity level even if they are older and rotate much
slower than more massive stars. As it was shown, the main parameter
controlling activity is the Rossby number, not rotation period. The
results of the present paper show that the empirical turnover time
increases with decreasing mass, at least for early M types. Rotation
periods of the old disk M type stars are not much longer than their
turnover times so the resulting Rossby numbers are of the order of
unity. Such stars should be considerably more active than massive stars
with the same rotation periods but shorter turnover times. Our analysis
applies to early M type stars up to about M4, with masses above the full
convection limit. There is indication that empirical turnover time
increases further for stars with masses as low as 0.17~\MS\ but a
significantly more numerous sample is needed to draw meaningful
conclusions.

\Section{Conclusions}
We checked 180 nearby X-ray active M type stars for periodic light
variations. Analysis of 6 seasons of ASAS photometric observations led
to the detection of periodic variations in 31 objects with typical
half-amplitude of 10~mmag and periods between a fraction of a day and
83~d. We interpret these variations as resulting from rotational
modulation of non-uniformly spotted stellar surface. Prior to our search
only three M type stars with long (at least a few tens of days)
rotation periods were known. We detected 9 more stars with rotation
periods longer than 10~d which is the approximate limiting period of
saturation for early M dwarf stars. Our data show that also M type stars
obey period-activity relation such that activity decreases with
increasing period. The relation is mass dependent, similarly as in G and
K type stars. To obtain a single period-activity relation rotation for
all stars, their periods should be scaled down by mass dependent factors
(turnover times). The analyzed stars were divided into four mass
intervals and turnover times were determined for three of them. The
fourth bin, containing the least massive stars has too few data points
to obtain its reliable value but a continuous increase of the turnover
time with decreasing mass, down to masses of 0.1--0.2~\MS, is suggested.
We demonstrated that the formula for spin down rate of single stars,
derived by Stępie\'n (1988) also correctly predicts the evolution of
rotation period of M type stars when newly determined turnover times are
used. The analysis of space motions showed that all OD stars have long
rotation periods with a median value of 47 days. This clearly
demonstrates that M type stars spin down as they age. Assuming 10~Gyr
for the age of OD stars we showed that the exponential period-activity
relation, together with the spin down rate formula, correctly predicts
the decrease of X-ray activity with age, as observed by Feigelson \etal
(2004). The observations of Feigelson \etal are at odds with an
age--activity power relation resulting from the Skumanich law and
period-activity relation suggested by Pallavicini \etal (1981) and
Pizzolato \etal (2003). 

It would be desirable to extend the period search to late M stars but to
this purpose photometric observations reaching fainter stars than
present day ASAS are needed.

\Acknow{This work was partly supported by MNiSW grant 1 P03 016 28. We
are particularly grateful to Dr Grzegorz Pojma\'nski for help in using
ASAS data and many useful comments. This research has made use of the
SIMBAD database, operated at CDS, Strasbourg, France.}


\begin{references}
\refitem{Argiroffi, C., Maggio, A., Peres, G., Stelzer, B., and Neuh\"auser, R.}{2005}{\AA}{439}{1149}
\refitem{Andersen, J.}{1991}{Astron. Astrophys. Rev.}{3}{91}
\refitem{Baraffe, I., Chabrier, G., Allard F., and Hauschildt, P.H.}{1998}{\AA}{337}{403}
\refitem{Barnes, J.R., and Collier Cameron, A.}{2001}{\MNRAS}{326}{950}
\refitem{Barnes, S.A.}{2003}{\ApJ}{586}{997}
\refitem{Barry, D.C.}{1988}{\ApJ}{334}{436} 
\refitem{Benedict, G., \etal}{1998}{\AJ}{116}{429}
\refitem{Beuzit, J.-L., \etal}{2004}{\AA}{425}{555} 
\refitem{Brandeker, A., Jayawardhana, R., and Najita, J.}{2003}{\AJ}{126}{2009} 
\refitem{Bopp, B.W., and Espenak, F.}{1977}{\AJ}{82}{916}
\refitem{Byrne, P.B., Doyle, J.G., and Butler, C.J.}{1984}{\MNRAS}{206}{907}
\refitem{Byrne, P.B., Black, E., and The, P.S.}{1987}{\AA}{186}{261} 
\refitem{Delfosse, X., Forveille, T., Perrier, C., and Mayor, M.}{1998}{\AA}{331}{581}
\refitem{Delfosse, X., Forveille, T., Ségransan, D., Beuzit, J.-L., Udry, S., Perrier, C., and Mayor, M.}{2000}{\AA}{364}{217} 
\refitem{Donahue, R.A., Saar, S.H., and Baliunas, S.L.}{1996}{\ApJ}{466}{384}
\refitem{Durney, B.R., and Latour, J.}{1978}{Geophys. Astrophys. Fluid Dyn.}{9}{241}
\refitem{Eggen, O.J.}{1969}{PASP}{81}{553}
\refitem{Evans, D.S.}{1959}{\MNRAS}{119}{526}
\refitem{Feigelson, E.D., Hornschemeier, A.E., Micela, G., Bauer, F.E., Alexander, D.M., Brandt, W.N., Favata, F., and Sciortino, S.}{2004}{\ApJ}{611}{1107}
\refitem{Fleming, T.A., Schmitt, J.H.M.M., and Giampapa, M.S.}{1995}{\ApJ}{450}{401}
\refitem{Franciosini, E., Randich, S., and Pallavicini, R.}{2003}{\AA}{405}{551}
\refitem{Gilman, P.}{1980}{~}{~}{in: ``Stellar Turbulence'', Eds. D.F. Gray and J.L. Linsky, Springer, Berlin, p.19}
\refitem{Gilliland, R.L.}{1985}{\ApJ}{299}{286}
\refitem{Gizis, J.E., Reid., I.N., and Hawley, S.L.}{2002}{\AJ}{123}{3356}
\refitem{Golimowski, D., \etal}{2004}{\AJ}{128}{1733}
\refitem{Gray, D.F.}{1982}{\ApJ}{261}{259}
\refitem{G\"udel, M., Guinan, E.F., and Skinner, S.L.}{1997}{\ApJ}{483}{947}
\refitem{Guinan, E.F., and Morgan, N.D.}{1996}{BAAS}{188}{No. 71.05}
\refitem{Haisch, B., and Schmitt, J.H.M.M.}{1996}{PASP}{108}{113}
\refitem{Hempelmann, A., Schmitt, J.H.M.M., Schultz, M., R\"udiger, G., and
Stępie{\'n}, K.}{1995}{\AA}{294}{515}
\refitem{Hartmann, L.W., and Noyes, R.W.}{1987}{Ann. Rev. Astron. Astrophys.}{25}{271}
\refitem{Hawley, S.L., Reid, I.N., and Tourtellot, J.G.}{2000}{~}{~}{in: ``Very Low Mass Stars and Brown Dwarfs'', Eds. R. Rebolo and M.R. Zapatero-Osorio, Cambridge Univ. Press, Cambridge, p.~109}
\refitem{Herbig, G.H., and Moorhead, J.M.}{1965}{\ApJ}{141}{649}
\refitem{H\"unsch, M., Schmitt, H.H.M.M., Sterzik, M.F., and Voges,
W.}{1999}{\AAS}{135}{319 (NEXUS)}
\refitem{Jao, W.-Ch., Henry, T.J., Subasavage, J.P., Bean, J.L., Costa, E.,
Ianna, P.A., and M\'endez, R.A.}{2003}{\AJ}{125}{332}
\refitem{J\"arvinen, S.P., Berdyugina, S.V., Tuominen, I., Cutispoto, G., and Bos, M.}{2005}{\AA}{432}{657}
\refitem{Jeffries, R.D., and Bromage, G.E.}{1993}{\MNRAS}{260}{132}
\refitem{Kawaler, S.D.}{1988}{\ApJ}{333}{236}
\refitem{Kim, Y., and Demarque, P.}{1996}{\ApJ}{457}{340}
\refitem{K\"urster, M., Schmitt, J.H.M.M., Cutispoto, G., and Dennerl, K.}{1997}{\AA}{320}{831}
\refitem{Maggio, A., \etal}{1987}{\ApJ}{315}{687}
\refitem{Messina, S., Pizzolato, N., Guinan, E.F., and Rodon\'o, M.}{2003}{\AA}{410}{671}
\refitem{Meusinger, H., Stecklum, B., and Reimann, H.-G.}{1991}{\AA}{245}{57}
\refitem{Mohanty, S., and Basri, G.}{2003}{\ApJ}{583}{472}
\refitem{Mokler, F., and Stelzer, B.}{2002}{\AA}{391}{1025}
\refitem{Montes, D., G\'alvez, Fernandez-Figueroa, M.J., and Crespo-Chac\'on I.}{2006}{Ap. Sp. Sci.}{304}{365}
\refitem{Noyes, R.W., Hartmann, L.W., Baliunas, S.L., Duncan, D.K., and Vaughan, A.H.}{1984}{\ApJ}{279}{763}
\refitem{Pace, G., and Pasquini, L.}{2004}{\AA}{426}{1021}
\refitem{Pallavicini, R., Golub, L., Rosner, R., Vaiana, G.S., Ayres, T., and Linsky, J.L.}{1981}{\ApJ}{248}{279}
\refitem{Pettersen, R.R.}{1983}{IAU Coll.}{71}{17}
\refitem{Pizzolato, N., Maggio, A., Micela, G., Sciortino, S., and Ventura, P.}{2003}{\AA}{397}{147}
\refitem{Pojma\'nski, G.}{1997}{\Acta}{47}{467}
\refitem{Pojma\'nski, G.}{2004}{Astron. Nach.}{325}{553}
\refitem{Reid, N., Hawley, S.L., and Mateo, M.}{1995}{\MNRAS}{272}{828}
\refitem{Rucinski, S.M., and VandenBerg, D.A.}{1986}{PASP}{98}{669}
\refitem{Schmitt, J.H.M.M., and Liefke, C.}{2004}{\AA}{417}{651 (NEXXUS database)}
\refitem{Scholz, A., and Eisl\"offel J.}{2005}{\AA}{429}{1007}
\refitem{Schwarzenberg-Czerny, A.}{1989}{MNRAS}{241}{153 (AoV)}
\refitem{Skumanich, A.}{1972}{\ApJ}{171}{565}
\refitem{Sills, A., Pinsonneault, M.H., and Terndrup, D.M.}{2000}{\ApJ}{534}{335}
\refitem{Silvestri, N.M., Hawley, S.L., and Oswalt, T.D.}{2005}{\ApJ}{129}{2428}
\refitem{Soderblom, D.R., Duncan, D.K., and Johnson, D.R.H.}{1991}{\ApJ}{375}{722}
\refitem{Stępie\'n, K.}{1988}{\ApJ}{335}{907}
\refitem{Stępie\'n, K.}{1989}{\AA}{210}{273}
\refitem{Stępie\'n, K.}{1994}{\AA}{292}{191}
\refitem{Stępie\'n, K.}{2003}{~}{~}{in: ``Modeling of Stellar Atmospheres'', {\it IAU Symp.} {\bf 210}, Eds. N. Piskunov, W.W. Weiss and D.E. Gray, ASP, p.~716}
\refitem{Stępie\'n, K.}{2006}{\Acta}{56}{199}
\refitem{Torres, C.A.O., Busco, I.C., and Quast, G.R.}{1983}{IAU Coll.}{71}{175}
\refitem{Wertheimer, J.G., and Laughlin, G.}{2006}{\AJ}{132}{1995}
\end{references}
\end{document}